# A Bayesian approach to strong lens finding in the era of wide-area surveys

Philip Holloway [1]★, Philip J. Marshall,[2,3] Aprajita Verma,[1] Anupreeta More[4,5] Raoul Cañameras,[6,7] Anton T. Jaelani,[8,9] Yuichiro Ishida[10] and Kenneth C. Wong[11,12]

[1]*Sub-department of Astrophysics, University of Oxford, Denys Wilkinson Building, Keble Road, Oxford OX1 3RH, UK*
[2]*Kavli Institute for Particle Astrophysics and Cosmology, Department of Physics, Stanford University, Stanford, CA 94305, USA*
[3]*SLAC National Accelerator Laboratory, Menlo Park, CA 94025, USA*
[4]*The Inter-University Centre for Astronomy and Astrophysics (IUCAA), Post Bag 4, Ganeshkhind, Pune 411007, India*
[5]*Kavli Institute for the Physics and Mathematics of the Universe (IPMU), 5-1-5 Kashiwanoha, Kashiwa-shi, Chiba 277-8583, Japan*
[6]*Max – Planck – Institut für Astrophysik, Karl – Schwarzschild – Str 1, D-85748 Garching, Germany*
[7]*Department of Physics, TUM School of Natural Sciences, Technical University of Munich, James – Franck – Str 1, D-85748 Garching, Germany*
[8]*Astronomy Research Group and Bosscha Observatory, FMIPA, Institut Teknologi Bandung, Jl. Ganesha 10, Bandung 40132, Indonesia*
[9]*U-CoE AI-VLB, Institut Teknologi Bandung, Jl. Ganesha 10, Bandung 40132, Indonesia*
[10]*Department of Earth and Planetary Sciences, School of Science, Kyushu University, 744 Motooka, Nishi-ku, Fukuoka 819-0395, Japan*
[11]*Research Center for the Early Universe, Graduate School of Science, The University of Tokyo, 7-3-1 Hongo, Bunkyo-ku, Tokyo 113-0033, Japan*
[12]*National Astronomical Observatory of Japan, 2-21-1 Osawa, Mitaka, Tokyo 181-8588, Japan*

**ABSTRACT**
The arrival of the Vera C. Rubin Observatory's Legacy Survey of Space and Time (LSST), Euclid-Wide and Roman wide-area sensitive surveys will herald a new era in strong lens science in which the number of strong lenses known is expected to rise from $\mathcal{O}(10^3)$ to $\mathcal{O}(10^5)$. However, current lens-finding methods still require time-consuming follow-up visual inspection by strong lens experts to remove false positives which is only set to increase with these surveys. In this work, we demonstrate a range of methods to produce calibrated probabilities to help determine the veracity of any given lens candidate. To do this we use the classifications from citizen science and multiple neural networks for galaxies selected from the Hyper Suprime-Cam survey. Our methodology is not restricted to particular classifier types and could be applied to any strong lens classifier which produces quantitative scores. Using these calibrated probabilities, we generate an ensemble classifier, combining citizen science, and neural network lens finders. We find such an ensemble can provide improved classification over the individual classifiers. We find a false-positive rate of $10^{-3}$ can be achieved with a completeness of 46 per cent, compared to 34 per cent for the best individual classifier. Given the large number of galaxy–galaxy strong lenses anticipated in LSST, such improvement would still produce significant numbers of false positives, in which case using calibrated probabilities will be essential for population analysis of large populations of lenses and to help prioritize candidates for follow-up.

**Key words:** gravitational lensing: strong – methods: data analysis – methods: statistical.

## 1 INTRODUCTION

Gravitational lensing is the deflection of light due to gravity as it passes a massive object. This can occur at a wide range of mass scales, ranging from microlensing of single stars by a compact source (e.g. a black hole; Sahu et al. 2022), to lensing of galaxies by galaxy clusters (e.g. Southern MAssive Cluster Survey (SMACS) 0723; Pascale et al. 2022). Strong lensing occurs when, from the perspective of the observer, the source appears as multiple images surrounding the lens. Strong lensing is a rare occurrence: only the most massive galaxies can deflect the source light sufficiently to be resolvable from the lensing galaxy. The number of galaxy-scale strong lenses identified to date is of order 1000 (hundreds of which have been confirmed, e.g. Bolton et al. 2008; Brownstein et al. 2012). These have been found across a range of surveys and telescopes, such as the Dark Energy Survey (DES, e.g. Jacobs et al. 2019a, b), Canada-France-Hawaii Telescope Legacy Survey (CFHTLS, Marshall et al. 2016; More et al. 2016; Jacobs et al. 2017), Canada-France Imaging Survey (CFIS, Savary et al. 2022), Sloan Digital Sky Survey (SDSS, e.g. Bolton et al. 2006, 2008), Hyper-Suprime Cam (Sonnenfeld et al. 2020; Cañameras et al. 2021; Shu et al. 2022), *Hubble Space Telescope* (Faure et al. 2008; Jackson 2008; Pawase et al. 2014), Kilo-Degree Survey (KiDS, Petrillo et al. 2017, 2019; He et al. 2020; Li et al. 2021) including in the radio (e.g. the Very Large Array; Browne et al. 2003; Myers et al. 2003). The known lenses therefore span a wide resolution range and all come with different selection functions. The number of lenses useful for a particular science case is typically much smaller than the number of known lenses, particularly if that science case requires spectroscopic redshifts or data at particular wavelengths. Such cases include constraining the stellar initial mass function over a lens population (Sonnenfeld et al. 2019) or the redshift evolution of the velocity dispersion function of early-type galaxies (ETG's, Geng et al. 2021) which both required spectroscopic redshifts, or the mass of the fuzzy-dark matter particle through high-resolution very long baseline interferometry observations (Powell et al. 2022, 2023).The ∼100 Sloan Lens ACS Survey (SLACS) lenses (Bolton et al. 2006, 2008) have been a go-to lens sample owing to the availability of redshift and high-resolution data. With the arrival of wide-area surveys such as Legacy Survey of Space and Time (LSST; Ivezić et al. 2019), the Euclid Wide Survey (Euclid Collaboration 2022), and Roman High Latitude Wide Area survey (Spergel et al. 2015; Akeson et al. 2019), strong lensing science will move from small samples of strong lenses to *population-level* analysis of the hundreds of thousands of lenses (Collett 2015; Holloway et al. 2023) identified by these surveys. Although the 4*MOST* Strong Lensing Spectroscopic Legacy Survey (Collett et al. 2023) will provide spectroscopic redshifts for ∼10 000 lens-source pairs, this will leave the majority without spectroscopic confirmation and follow-up visual inspection of the ∼100 000 strong lens candidates (and many more false positives) put forward by lens finding methods will be an unenviable task.

★ E-mail: philip.holloway@physics.ox.ac.uk

Strong lenses have been found using a wide variety of methods. These include machine learning [typically convolutional neural networks (CNN's), e.g. Lanusse et al. 2018; Schaefer et al. 2018 though other methods including Support Vector Machines (e.g. Hartley et al. 2017) and self-attention-based encoding models (Thuruthipilly et al. 2022) have also been used], citizen science (Geach et al. 2015; Marshall et al. 2016; More et al. 2016; Sonnenfeld et al. 2020; Garvin et al. 2022), arc-finding algorithms (e.g YATTALENS, Sonnenfeld et al. 2018; Arcfinder, Seidel & Bartelmann 2007), and expert visual inspection (Faure et al. 2008; Jackson 2008; Pawase et al. 2014). Although the automated methods e.g. CNN's are continually improving over time, they still require visual follow-up inspection of high-scoring candidates by strong lens experts to remove false positives. Even with small-medium sized surveys, this is time consuming and difficult: strong lens identification is not clear-cut, particularly with ground-based imaging or when using single-band imaging. From recent lens searches (Sonnenfeld et al. 2020; Cañameras et al. 2021; Rojas et al. 2022) the number of high-scoring candidates which underwent visual inspection exceeded the resulting number of A–B grade lenses by factors ≥10×. Furthermore, Rojas et al. (2023) found that ∼6 expert graders are required to reduce the classification error below 0.1 (on a 0–1 grading system). With wide-area surveys, lens finding methods which scale-up easily will be key to identifying most lenses; however, these will still leave a large number of high-scoring false positives to remove. This paper aims to address this problem. The goals of this paper are two-fold:

(i) **To produce calibrated probabilities that a given galaxy system is a lens**. We define a calibrated score X as one for which systems with this score are indeed lenses X per cent of the time, as verified on a distinct test set. For wide-area surveys, follow-up spectroscopy or visual inspection of all lens candidates will not be possible. It will therefore be important to have accurate probabilities for lens candidates in order to perform unbiased population-level analysis, the alternative being significantly reducing the sample size to those which have been spectroscopically or visually confirmed. Furthermore, having accurate probabilities allows direct comparison of lens candidates across different lens finders.

(ii) **To create an ensemble strong lens classifier**. A given survey may be targeted by multiple lens finding methods, each with their own strengths and weaknesses. Neural network ensembles have been investigated previously (e.g. Schaefer et al. 2018; Canameras et al. 2023 for galaxy–galaxy lensing and Andika et al. 2023 for lensed quasars), simply averaging over the individual network scores. We wish to take this method further, by incorporating the calibrated probabilities generated above in a Bayesian framework, as well as including a citizen science classifier to diversify our ensemble beyond neural networks.

In the process of reaching these goals, we aim to answer the following questions:

(i) Is there a significant improvement in performance using an ensemble method compared to a single classifier?

(ii) Are there regions of parameter space which suit certain methods best?

(iii) How does the degree of improvement change when combining only neural network classifiers, compared to combining neural networks with a citizen science classifier?

This paper is structured as follows: We detail the data used in this work in Section 2. In Sections 3.1 and 3.2, we summarize and apply different classifier calibration methods. In Section 3.3, we then combine these calibrated probabilities into a single ensemble classifier. Our results are presented in Section 4 and discussed further in Section 5, including implications for forthcoming lens searches in LSST and other wide-area surveys in Section 5.4. We conclude in Section 6.

## 2 DATA

To develop our ensemble classifier, we used the classification outputs from six strong lens finders applied to Hyper-Suprime Cam (HSC) Subaru Strategic Program data (Aihara et al. 2022). This, being a ground-based survey over hundreds of square degrees, was a useful proxy for LSST-like data. The wide-area survey is being conducted in *grizy* bands, covering ∼ 1200 square degrees. The HSC S17A and Public Data Release 2 (PDR2) data releases (Aihara et al. 2018, 2019) covered 1026 square degrees (225) and 1114 square degrees (305), respectively, in at least one band (all bands) to a $5\sigma$ depth of $i$ ∼ 26.2. Only the *gri* bands were used in this work. The strong lens classifiers used are described below:

(i) **Citizen Science** (Sonnenfeld et al. 2020): A citizen science search by the Space Warps project, which used the HSC S17A data release (Aihara et al. 2018, 2019) and provided citizen classifications for ∼300 000 objects. The galaxies selected for this search had photometric redshifts $0.2 < z < 1.2$ and inferred stellar masses $\log(M_*/M_\odot) > 11.2$.

(ii) **Neural Network 1** (Cañameras et al. 2021): A ResNet classifier, which used HSC PDR2 data (Aihara et al. 2019) over a much larger sample of objects (network scores and positions were available for $5.4 \times 10^7$ objects). The objects chosen for this search were required to have *i*-band Kron radius ≥0.8 arcsec, to narrow the search to Luminous Red Galaxies with image separations likely greater than the median seeing.

(iii) **Neural Network 2** (Shu et al. 2022): A ResNet classifier from Lanusse et al. (2018), using HSC PDR2 data and in particular targeting high-redshift lens systems. The primary selections for this search were colour cuts: $0.6 < g - r < 3.0$ and $2.0 < g - i < 5.0$ to identify red, high-redshift galaxies. These cuts were taken from Jacobs et al. (2019b) to select lens galaxies with $z \geq 0.8$. The lenses in the training set of this network lay in the range $0.1 \lesssim z \lesssim 1.0$.

(iv) **Neural Network 3**: The second classifier presented by Shu et al. (2022), differing from the first in having a higher fraction of z

> 0.6 lenses in the training set, and image pre-processing. The lenses in the training set of this network lay in the range $0.4 \leq z \leq 1$.

(v) **Neural Networks 4** (Ishida et al., in prep.) **and 5** (Jaelani et al. 2023): These networks were both convolutional neural networks applied to HSC PDR2. They shared the same training set, in particular targeting smaller Einstein radii >0.5 arcsec (the distribution of Einstein radii in the training distribution of Einstein radii decreased roughly exponentially from 0.5 to 3.0 arcsec while Networks 1–3 focused on $\theta_E \gtrsim 0.75$ arcsec). The galaxies were originally selected based on pre-selection criteria, such as stellar mass $> 5 \times 10^{10}\,M_\odot$, specific star formation rate $< 10^{-10}$, and redshift range $0.2 < z < 1.2$. However, in this work, these CNNs were applied to the sample of galaxies with classifier outputs available from all of the first four classifiers listed.

The objects in each catalogue were cross-matched, with a maximum separation of 1 arcsec, to produce a sample of 126 312 galaxies with both citizen science and neural network outputs. The main cuts from the parent samples which affected the resulting cross-matched sample were therefore:

(i) Photometric redshift: $0.2 < z < 1.2$ (CS)
(ii) Stellar mass: $\log(M_*/M_\odot) > 11.2$ (CS)
(iii) Kron radius (*i*-band): $R_{\mathrm{Kron},i} \geq 0.8$ arcsec (NN 1)
(iv) Colour cut 1: $0.6 < g - r < 3.0$ (NN 2 + 3)
(v) Colour cut 2: $2.0 < g - i < 5.0$ (NN 2 + 3)

A subset (109 128 objects), with ≥5 citizen classifications, was used in the following analysis. On investigating the properties of the cross-matched and parent samples, we found the properties of the lens candidates (such as redshift, stellar mass, or colour) found by each classifier were similar even though the parent distributions varied between classifiers. It was therefore reasonable to combine these classifiers into an ensemble. In order to calibrate the output scores of the classifiers, we required a 'ground truth', i.e. a list of classified objects of known lenses/non-lenses. We collated such a list using expert grades from four sources: the Survey of Gravitationally-lensed Objects in HSC Imaging VI (SuGOHI VI, Sonnenfeld et al. 2020), Highly Optimized Lensing Investigations of Supernovae, Microlensing Objects, and Kinematics of Ellipticals and Spirals VI (HOLISMOKES VI, Cañameras et al. 2021), HOLISMOKES VIII (Shu et al. 2022), and the online SuGOHI data base[1] of lens candidates. We then assigned each cross-matched object with an overall grade according to the following order: HOLISMOKES VI, SuGOHI VI, SuGOHI data base, and HOLISMOKES VIII. These grades (G) were indications of the confidence that a given object was a lens. They were defined in the range [0,3] where $G = 3$ indicated 'certain lens' (e.g. multiple images, in the correct configuration), $G = 2$ indicated 'probable lens' (most features consistent with a strong lens), and $G = 1$ indicated 'possible lens' (e.g. a single arc, which could well be a contaminant). In this work, we refer to A–B grade lenses as those with $G \geq 1.5$, and A–C grade lenses as those with $G \geq 1.0$. There were 34 objects whose grades differed by ≥2 which were inspected by PH and re-assigned an appropriate grade (typically taking the average of the relevant grades except in clear-cut cases). In total, 3744 cross-matched objects had assigned grades, of which 189 were graded A–B and were treated as true lenses.

[1] http://www-utap.phys.s.u-tokyo.ac.jp/~oguri/sugohi/

## 3 METHOD

Our first aim was to produce calibrated scores for each classifier individually. The graded objects were split into three sets: a training ('calibration') set, a validation set and a test set, in ratios 2:1:1. For this work, we considered A+B grade lenses as true lenses, and the remainder (graded or otherwise) as non-lenses. Excluding the ungraded objects would lead to unrealistic values for the receiver operating characteristic (ROC) curve and purity-completeness summary statistics and it is likely the vast majority of true lenses were identified by at least one of our sources of expert grades.

### 3.1 Summary of calibration methods

The purpose of calibration is to produce accurate probabilities that a given object is a lens. The true distribution of (raw) classifier scores which a large population of lenses (and non-lenses) would receive from a given classifier is generally not known but could be inferred from a sample of data (this is known as density estimation). The calibration methods described below are similar to such density estimation, however, here we produce a continuous distribution function of the proportion of objects with a given score which are lenses, inferred from a finite sample of classifier scores and their classifications. These are then tested against a separate 'test' set to verify the calibration is robust.

We demonstrate the various common methods available to map a classifier score to a probability via a toy model shown in Fig. 1. For this, $10^4$ 'systems' were generated, with classifier scores ($x$) assigned uniformly in the range [0,1] and a class (lens or not) which was allocated randomly with probability $P_{\mathrm{true}}$ given by a quartic function of its score, $P_{\mathrm{true}}(x)$; this function is shown in Fig. 1(a). This function represents 'perfect' calibration as for each score it maps to the true probability a given system is a lens. The calibration methods considered in this work are described below with their calibration curves for this toy data shown in the remaining panels of this figure.

(i) Isotonic regression (Zadrozny & Elkan 2002) is a non-parametric, discontinuous fit of a monotonically increasing curve to the classifier output distribution.

(ii) Platt scaling (Platt 2000) is a parametric calibration method of the form:

$$P(L|x) = \frac{1}{1 + \exp(Ax + B)}, \quad (1)$$

where $A$ and $B$ are fitted to the data and $P(L|x)$ refers to the probability that a system is a lens (=L) given a classifier score $x$. Although commonly used, this requires the calibration curve to closely match a sigmoid function, which *a priori* may not be true.

(iii) Kullback Leibler importance estimation procedure (KLIEP; Sugiyama et al. 2008): This minimizes the KL divergence of the ratio $f(x|L)/f(x)$ with probability density functions $f(x|L)$ and $f(x)$ of a classifier score $x$ (given the subject is a lens in the first case). From Bayes' Theorem, weighting this by $P(L)$ gives the probability a subject is a lens, given a classifier score. KLIEP uses a Gaussian mixture model, with Gaussians of fixed width placed at the positions of the lenses in classifier-output space. Their respective weights are tuned to minimize the KL divergence.

(iv) Variable bin fitting: This was designed to account for the small number of lenses in our sample, compared to the much larger number of non-lenses. The method was adapted from that for generating a histogram but here we allowed the bins to overlap and rather than using bins of fixed width, they had a fixed number of *lenses* in each (and thus a varying total bin occupancy). We chose to use bins of five lenses. Bins at lower scores were therefore wider (due to the lower occurrence of lenses) while at higher scores were narrower. The calibration curve was then formed from the fraction of lenses in each bin compared to the total bin occupation. This provided higher resolution calibration in regions where there were lots of lenses (i.e. for high classifier scores), while averaging out the calibration where lenses were sparse (i.e. for low classifier scores). Fig. 1(d) shows a demonstration of this on toy data with a subset of the bins shown.

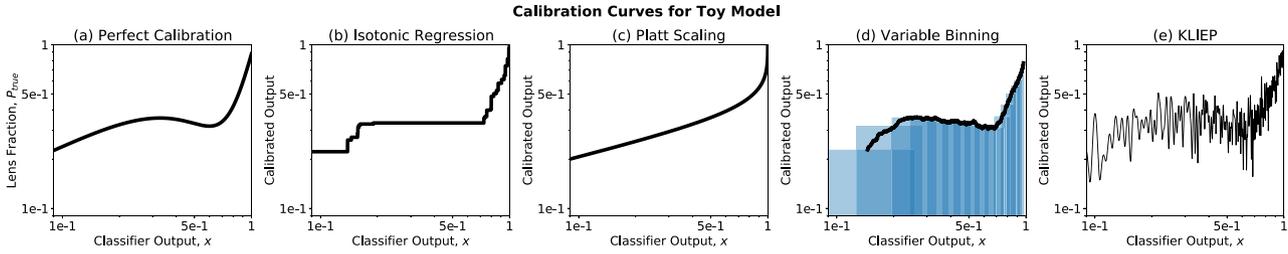

**Figure 1.** Calibration curves applied to a toy model for the range of calibration methods considered. The perfect calibration curve is shown on the far left. Platt scaling is the worst performing as the class distribution does not follow a sigmoid shape and thus is not considered further in this work. A subset of the bins used to generate the calibration curve in (d) are shown; the bin widths become smaller for larger classifier outputs as the fraction of lenses increases.

### 3.2 Application of calibration methods

We applied a subset of calibration methods to our data. We did not apply Platt Scaling as we did not expect our calibration curves to match a sigmoid function. A key feature of the data was the significant class imbalance: only ∼ 0.2 per cent of all objects used in this work were lenses, and these are more concentrated at high classifier scores. We used this to achieve higher-resolution calibration for high-scoring subjects.

For each calibration method, we applied the calibration algorithm to the rank values of the classifier outputs, rather than the outputs themselves. The ranks are defined as the ordered positions of the systems from lowest to highest raw classifier output for a given classifier; they do not depend on expert grade. We then interpolated back from rank to classifier score to produce the calibration map. We wished our calibration method to be independent of the type of classifier used and the score distribution of its output. The distribution of outputs can change significantly between different classifiers acting on the same objects, however an excess of high (raw) scoring objects from a particular classifier should not be treated as all of these objects having high calibrated probabilities. Using the ranks removed this effect and simply assumes that a higher classifier score (qualitatively) implies higher confidence of a lens.

The KLIEP and variable binning methods have hyperparameters we fixed prior to calibration. We used a Gaussian kernel of width $\sigma$ = 40; a balance between overfitting to the training data (occurring from a smaller kernel) and preventing oversmoothing at high scores (from a larger kernel). The KLIEP Gaussian mixture model had the same number of kernels as lenses. As shown by the calibration curve, the fraction of objects which are lenses increases significantly for high scores: using the rank values allowed finer tuning to these regions while not overfitting at lower classifier scores. Fig. 2 shows the calibration curves for each of the methods applied. In all cases the curves steepen for high scores: choice of score threshold will have a significant impact on purity in this region. The calibration curves do not follow the $y = x$ line (dotted), indicating that (as expected) the original classifier outputs were not already calibrated. The mappings from the variable bin and isotonic regression method are relatively similar to each other, across all classifiers; the KLIEP method differs more significantly as the probability density function (PDF) only becomes non-zero close to the positions of the lenses in the training set. This effect would be reduced by increasing the kernel size, but would lead to underfitting at the highest classifier scores. Since we were primarily focused on high-probability candidates, we prioritized reducing this underfitting.

We validated these calibration curves against a separate test set, shown in Fig. 3. These show that spanning a wide range on log-$p$, the calibration mapping can produce accurate probabilities. The isotonic calibration produced the smallest variation upon bootstrapping, so is used for the subsequent analysis.

### 3.3 Summary of combination methods

Given now-calibrated classifier outputs, we considered methods to combine them to produce a single score. The combined score will not necessarily itself be calibrated, thus further calibration stage using the methods above may be required. The aim of this classifier combination was to maximize the purity of the resultant sample. We tested three different methods: a generalized mean, dependent Bayesian combination, and independent Bayesian combination. The latter two methods produce a posterior probability that a given object is a lens, given the calibrated outputs of the individual classifiers. The former is a simple *ad hoc* method for combining multiple scores but does not strictly produce a posterior probability.

#### 3.3.1 Generalized mean

We first considered a simple generalized mean of the form:

$$P_{\text{combo}} = \left( \frac{1}{N} \sum_{i=1}^{N} p_i^{\alpha} \right)^{\frac{1}{\alpha}}, \qquad (2)$$

where $N$ denotes the total number of classifiers in the ensemble, and $p_i$ is the calibrated probability for a particular object from the $i$th classifier. This takes on a variety of useful functions for different values of $\alpha$, in particular the arithmetic mean, harmonic mean, minimum and maximum of $p_i$ across the $N$ classifiers for $\alpha = 1$, $-1$, $-\infty$, and $+\infty$, respectively.

#### 3.3.2 Dependent Bayesian classifier combination

Although we have generated calibrated probabilities for each classifier, the dependence (agreement) of each classifier on another has not yet been quantified. For example if two identical classifiers both gave probabilities of 0.9 for the same object, our combined posterior should be 0.9 (the second, identical, classifier adds no new information), but if those classifiers were independent, we would expect the posterior would be >0.9. We here outline a Bayesian approach to model this dependence as follows. We calculate the posterior $P(L|\{R_i\})$, where $\{R_i\} \equiv \{R_1,...R_N\}$ is the set of score rankings for a given object from each classifier and the ground truth is labelled $L$ and $\bar{L}$ for 'lens' and 'non-lens', respectively. We denote the corresponding set of calibrated classifier scores as $\{C_i\}$. From Bayes' Theorem we can determine this posterior probability that a given system is a lens, as a function of the score rankings from all $N$ classifiers:

$$P(L|\{R_i\}) = \frac{f(\{R_2,...R_N\}|L, R_1)P(L|R_1)}{f(\{R_2,...R_N\}|R_1)}, \quad (3)$$

where $f$ denotes a multidimensional probability density function, $P$ denotes the probability of a discrete random variable, and $N$ denotes the total number of classifiers in the ensemble. $P(L|R_1)$ is known already, given a 1–1 relation between rank and calibrated score (i.e. the calibration mapping is strictly monotonic), as it is given by the calibrated output: $P(L|R_1) = C_1$. The choice of which classifier is denoted as $C_1$ is free, and as we will subsequently show, does not affect the result. We rewrite equation (3), in terms of the difference between the classifier outputs, with respect to classifier $C_1$: $\{\Delta_i\} = \{R_i\} - R_1$:

$$P(L|\{R_i\}) = \frac{f(\{\Delta_2,...\Delta_N\}|L, R_1)}{f(\{\Delta_2,...,\Delta_N\}|R_1)} \cdot C_1. \quad (4)$$

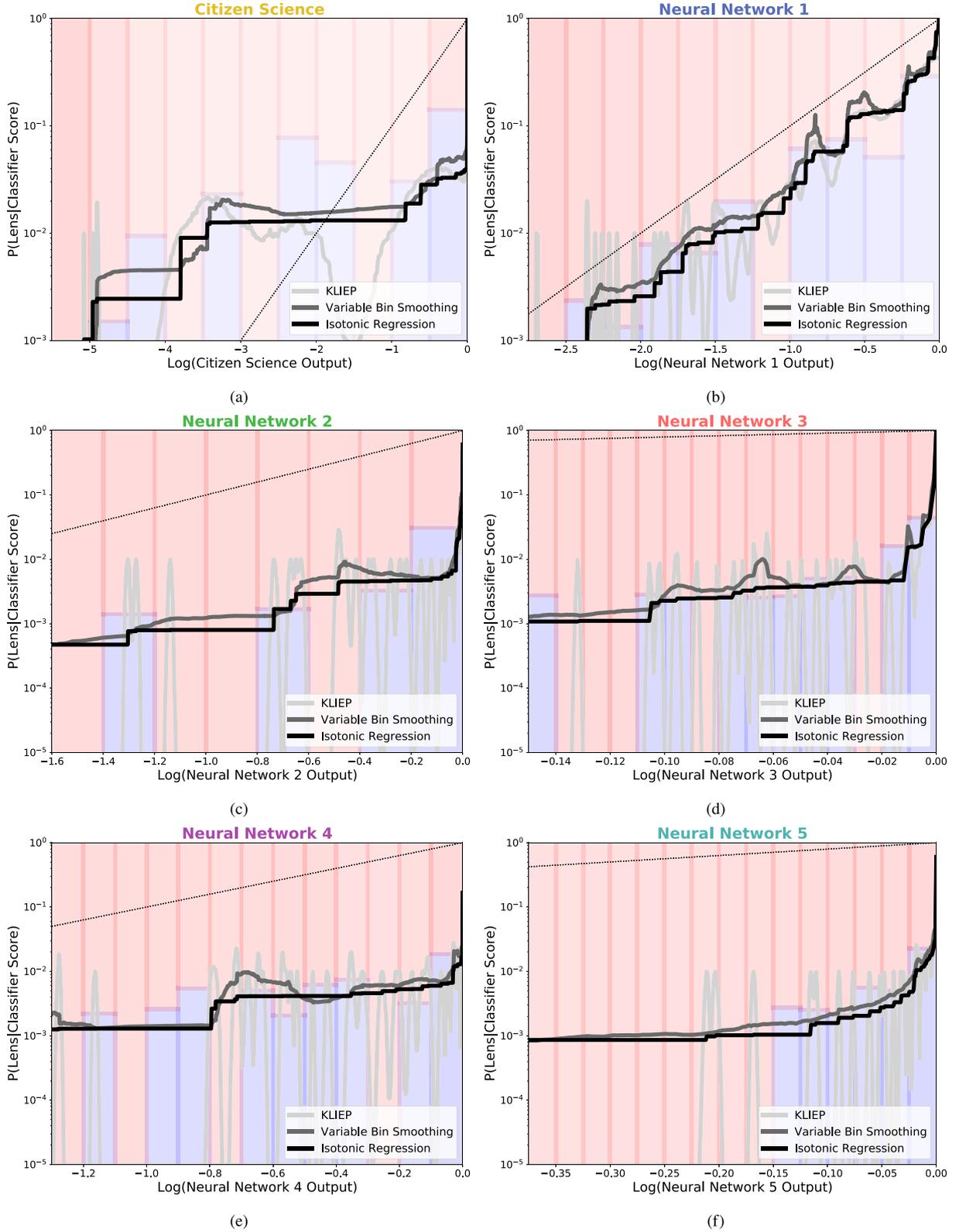

**Figure 2.** Calibration mapping for a range of calibration methods for each of the classifiers used in this work. For clarity, only the top 20 per cent of classifier scores are shown in each panel. The red and blue histograms refer to the fraction of lenses (blue/bottom) and non-lenses (red/top) in each bin. The mappings should be expected to roughly follow these histograms but not always, for example at the highest classifier scores, where the fraction of true lenses increases rapidly within a single bin. The bars are shaded by the number of objects in each bin. The dashed line is the $y = x$ line, which the calibration curves would follow if the outputs of the classifiers were already calibrated.

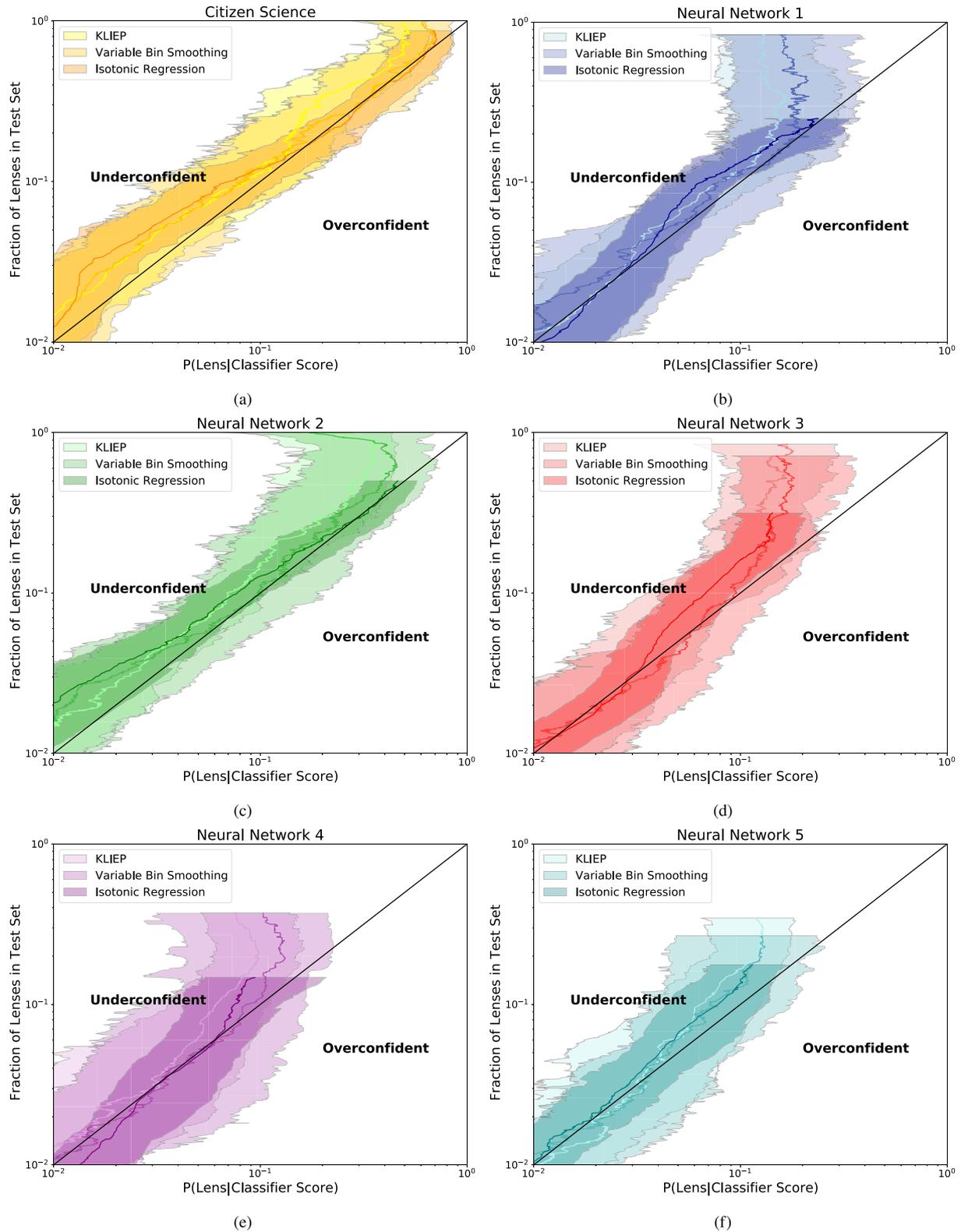

**Figure 3.** Validation of the calibration curves, applied to a separate test set of graded images. The $y$-axis is determined using the variable-binning method, and the error-bars are determined from bootstrapping. Each method shows good calibration (perfect calibration would be along the $y = x$ line), but are occasionally underconfident particularly for the highest scores.

We now model both numerator and denominator PDFs as Gaussian distributions, constant with respect to $R_1$. For independent classifiers, the denominator in equation (4) tends towards a triangle distribution (see Fig. 4); the distribution of $X-Y$, where $X$ and $Y$ are two random variables (i.e. the ranks) drawn from a uniform distribution, is a triangle distribution. If the classifiers are not completely independent, the distribution will deviate from a triangle distribution however both are well approximated by a Gaussian. We therefore used output ranking, rather than the classifier output itself as this provides a better fit to this distribution but this does not change the result. We now have:

$$P(L|\{R_i\}) = \frac{n(\{\Delta_2,...\Delta_N\}|\boldsymbol{\mu_{lens}}, \boldsymbol{\Sigma_{lens}})}{n(\{\Delta_2,...\Delta_N\}|\boldsymbol{\mu_{full}}, \boldsymbol{\Sigma_{full}})} \cdot C_1, \quad (5)$$

where $n$ denotes the Gaussian function. For six classifiers, the multidimensional Gaussians in equation (5) are five-dimensional. These multivariate normal distributions are independent to permutation of classifiers (i.e. which classifier is chosen to be $C_1$ does not change the best-fitting Gaussian distribution). However, this would change the value of $C_1$ (assuming all classifiers are not in exact agreement). With no reason, *a priori*, to favour one classifier over another, it is reasonable to average over the classifiers. Equation (5) becomes:

$$P(L|\{R_i\}) = \left(\frac{n(\{\Delta_2, \Delta_3, \Delta_4, \Delta_5, \Delta_6\}|\boldsymbol{\mu_{lens}}, \boldsymbol{\Sigma_{lens}})}{n(\{\Delta_2, \Delta_3, \Delta_4, \Delta_5, \Delta_6\}|\boldsymbol{\mu_{full}}, \boldsymbol{\Sigma_{full}})}\right) \cdot \langle C_i \rangle. \quad (6)$$

Note, this stems from our (strong) assumption that the ratio of PDFs in equation (4) can be modelled as a ratio of two fixed multivariate Gaussians; more complex functions would not suffer this problem. More flexible models (2 and 3-component mixture models of multivariate Gaussian distributions) were also tested, but did not provide significant improvements in the resultant combined calibrated score. The cases where the bracketed term in equation (6) exceeds 1 refer to when there is greater agreement between the individual classifiers than would be expected from the overall distribution (i.e. when the blue curve exceeds the black in Fig. 4) so the subject should therefore belong to the lens class.

### 3.3.3 Independent Bayesian classifier combination

We also considered a case where the results from each classifier were entirely independent of each other. From Bayes' Theorem, for a single classifier:

$$P(L|C_1) = \frac{f(C_1|L) \cdot P(L)}{f(C_1)} = \frac{f(C_1|L) \cdot P_0}{f(C_1|L) \cdot P_0 + f(C_1|\bar{L}) \cdot (1 - P_0)}. \quad (7)$$

For a set of calibrated probabilities, $\{C_i\}$, for a given object:

$$P(L|\{C_i\}) = \frac{P_0 \cdot \prod_{i=1}^{N} f(C_i|L)}{P_0 \cdot \prod_{i=1}^{N} f(C_i|L) + (1 - P_0) \cdot \prod_{i=1}^{N} f(C_i|\bar{L})}. \quad (8)$$

Since $\{C_i\}$ are in fact calibrated probabilities, we know:

$$C_i = \frac{N_L \cdot f(C_i|L)}{(N_L + N_{NL}) \cdot f(C_i)}, \quad (9)$$

where $N_L$ and $N_{NL}$ refer to the number of lenses and non-lenses in the (training) sample. We can therefore simplify equation (8):

$$P(L|\{C_i\}) = \frac{P_0 \cdot \prod_{i=1}^{N} \frac{C_i}{N_L}}{P_0 \cdot \prod_{i=1}^{N} \frac{C_i}{N_L} + (1-P_0) \cdot \prod_{i=1}^{N} \frac{(1-C_i)}{N_{NL}}}. \quad (10)$$

For an accurate prior: $P_0 = N_L/(N_L + N_{NL})$:

$$P(L|\{C_i\}) = \frac{N_L^{1-N} \cdot \prod_{i=1}^{N} C_i}{\left(N_L^{1-N} \cdot \prod_{i=1}^{N} C_i\right) + \left(N_{NL}^{1-N} \cdot \prod_{i=1}^{N}(1-C_i)\right)}, \quad (11)$$

where $N$ denotes the number of classifiers in the ensemble.

## 4 RESULTS

### 4.1 Testing the Bayesian combination approaches

We generated a toy model using simulated classifier scores to test the differences between the dependent and independent combination methods described above. A set of six 'independent' classifiers was generated as follows:

(a) A simulated sample of classifier scores was drawn from the distribution $y = 2x$ (for $x \in [0, 1]$). These scores were assigned as belonging to true lenses. A second, equally sized sample of scores was drawn from a $y = 2(1 - x)$ distribution, and assigned as non-lenses. By construction, the combined sample of scores was calibrated as we used an equal number of lenses and non-lenses.

(b) This was repeated for $N$ classifiers ($N = 6$ in this case).

Sets of six 'dependent' classifiers were generated as follows:

(c) A set of scores from distributions $y = 2x$ and $y = 2(1 - x)$ were drawn, as in stage 1) above. This was defined as Classifier 1.

(d) Five further classifiers were generated by adding varying degrees of random noise to the Classifier 1 scores. Varying the level of noise altered the degree of dependence (agreement) between classifiers.

e) Each classifier was then calibrated via isotonic regression.

We then compared the Receiver operating characteristic (ROC) curves measured using the dependent and independent Bayesian methods. We found that in the independent case, the Independent Bayesian classifier combination provided the best ensemble (in Area under ROC (AUROC) and completeness at false positive rate (FPR) of $10^{-3}$), while for the dependent case, the best combination method (Dependent or Independent) depended on the degree of noise added in Step d) above. For sets of real classifiers, we would therefore recommend applying both to verify which performs best.

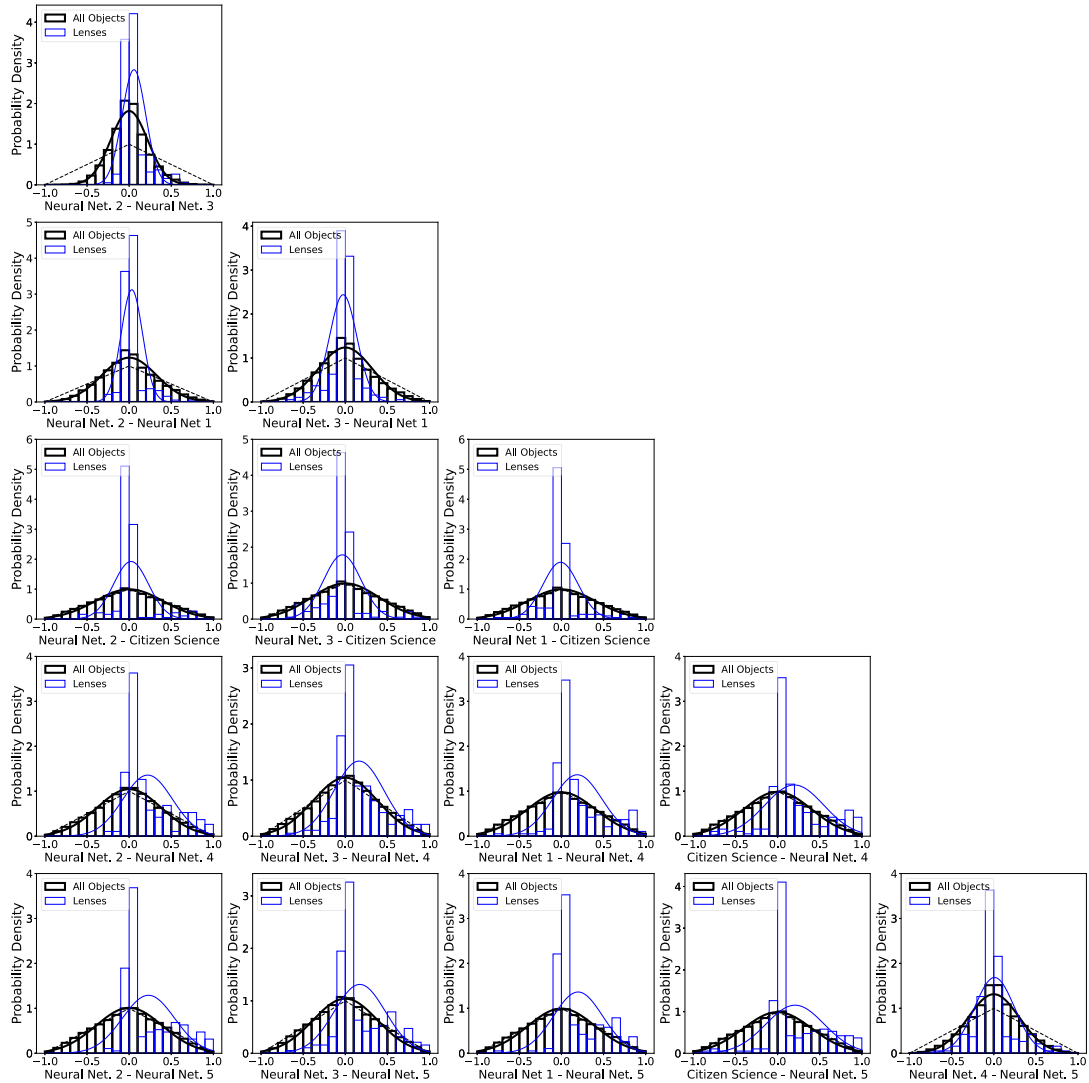

**Figure 4.** Distributions of the difference in ranking of the classifier scores for the cross-matched objects. The ranks were first normalized to take values in the range [0,1]. The black (bold) and blue histograms show the overall and lens distributions, respectively. The respective curves show the projections of the best-fitting multivariate Gaussian to the data. The dashed curve shows a triangle distribution (which hasn't been fitted to the data) which results from taking the difference between two uniformly distributed variables.

### 4.2 Applying the Bayesian combination methods

We investigated the dependence of the six classifiers used in this analysis. Fig. 4 shows the binned distributions of the difference in classifier ranking (with rankings normalized to 1), along with a multivariate Gaussian fit. The greatest correlation (for the whole cross-matched sample) is seen between the two HOLISMOKES VIII neural network classifiers; this is perhaps expected as these networks have the same architecture and similar non-lens training data. The distributions (black) shown in Fig. 4 which include the citizen science classifier (i.e. 'Neural Net. X – Citizen Science') are near-triangular. This demonstrates these rankings are nearly uncorrelated and suggests the networks and citizen classifiers found different objects easier/more difficult to classify (otherwise, the same objects would have received similar rankings from each). There was much greater agreement between classifiers when presented with a true lens (shown in blue in Fig. 4), than there was with non-lenses. It is this property which is used by the dependent Bayesian classifier combination method described in Section 3.3.2.

Upon inspecting the ensemble probabilities for a small number of spectroscopically confirmed lenses, we observed that in some cases, while the citizen science classifier would correctly identify the lens, the scores from the networks would not be sufficiently high to map to high probabilities. The networks could then effectively 'outvote' the citizen science classifier in the independent Bayesian ensemble. We therefore tested generating a network-only ensemble, recalibrating this, then further combining this ensemble with the citizen science classifier. We show the ROC curve for this method as 'Network Ensemble + CS' in Fig. 5 which shows the ROC curves for the individual classifiers along with the best-performing ensemble methods. A wider comparison of the ensemble methods described in Section 3.3 is given in Appendix A1. When considering the A–B lenses (Fig. 5a), the ensemble methods show improved classification over their individual constituent classifiers. We find 46 per cent completeness can be achieved with a false-positive rate of $10^{-3}$; by comparison, the best individual classifier achieved 34 per cent completeness on the same data set. How this relates to the sample purity is discussed in Section 5.4.

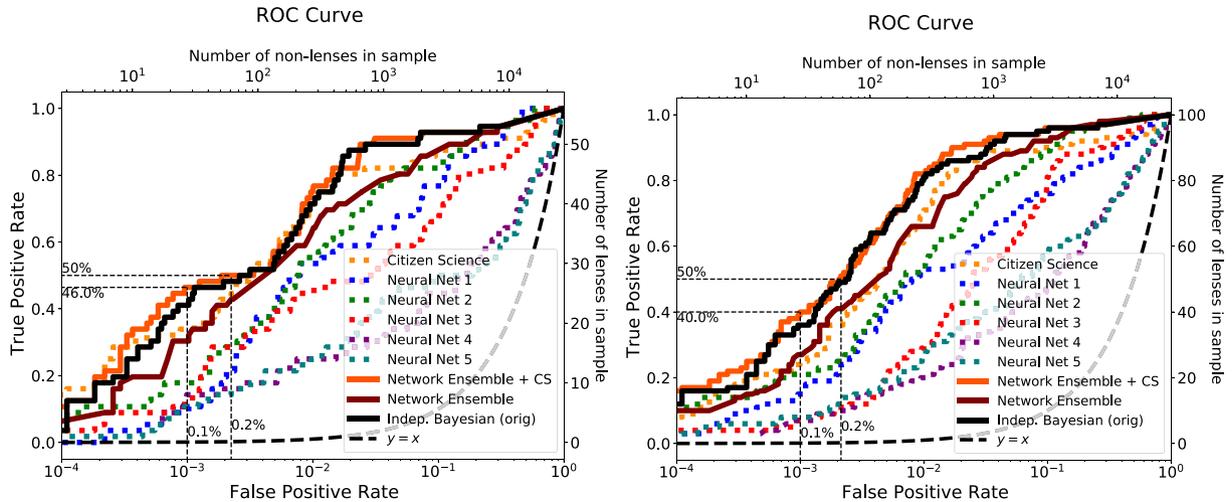

(a) Including A-B grade lenses as 'true lenses', without excluding cluster-scale candidates.

(b) Including A-C grade lenses as 'true lenses' and excluding cluster-scale candidates.

**Figure 5.** Receiver operating characteristic (ROC) curve for the individual lens classifiers (dashed) and combined methods, applied to a separate test set to that which the calibration methods were tuned. Ungraded subjects were treated as non-lenses. The dashed guidelines show the 50 per cent completeness and 0.1 per cent false-positive rate. The ROC curve for a random classifier is shown as the dashed curve ($y = x$). Note the $y$-axes are linear, while the $x$-axes are logarithmic.

## 5 DISCUSSION

### 5.1 Comparison with previous work

The overlap in lenses found by machine learning, citizen science, and spectroscopy was investigated by Knabel et al. (2020). They found very little overlap between the three methods: out of 107 lenses identified, only two were identified by more than one method (machine learning (ML) + citizen science (CS)). They attributed these two differences in the parent sample (e.g. different redshift cuts) and particular behaviours of each method (e.g. ML typically finding lenses similar to its training set). Our results are derived from a different parent sample but show citizen science and machine learning can be in much greater agreement. There are however two significant differences between our methods: first, Knabel et al. (2020) employed GalaxyZoo (Lintott et al. 2008; Marshall et al. 2016; Holwerda et al. 2019; Kelvin et al. in prep) as their citizen science classifier, which uses a question tree to identify the overall galaxy morphology, including the presence of lensing whereas Sonnenfeld et al. (2020) only looked for strong lenses. Furthermore, in this work we only compare cross-matched objects which both techniques classified as opposed to objects simply in the same field, removing the effect of differing selection functions for each sample. The difference in our results highlights the importance of object selection when conducting lens searches; a narrow selection could significantly reduce the number of lenses identified.

### 5.2 Comparison of citizen science versus a network ensemble

We investigated the qualitative properties of the galaxy systems identified and rejected by the citizen science classifier compared to those receiving high/low scores from an ensemble of five neural networks. Fig. 6 shows a selection of such cutouts. The upper (lower) quadrants show subjects which received a high- (low-) posterior probability from the network-only ensemble. The right- (left-) hand quadrants show subjects which received high- (low-) calibrated probabilities from the citizen science search. The vast majority of those which received very high (low) probabilities from both the ensemble and citizen science were correctly identified as (non) lenses. Those which received high-citizen scores but low-ensemble scores contained a mix of true lenses and interlopers (according to the expert grades used). There is no clear trend in these objects, but the presence of many bright bulges suggest that the network classifiers may have learnt to reject these; it should be noted that the networks did not have access to lens-subtracted images (unlike the citizen scientists) which would make some images, for example those with bright bulges but small Einstein radii, harder for the networks to classify. Furthermore, a small number of candidates were groups or had lensing features outside of the cutouts provided to the network, so it is unsurprising the networks rejected these – we tested the effect of this below (Section 5.3). The top-left quadrant of Fig. 6 shows a selection of objects which received high-ensemble probabilities but low-citizen science probabilities. These contained a number of face-on spiral galaxies where some networks may have been misidentified the spiral arms as lensed arcs.

### 5.3 Effect of ground-truth selection on classifier performance

As demonstrated in Section 5.2, some of the objects which the network ensemble assigned a low probability to were galaxy clusters, which would not have been included in their training sets. These would, however, have been identifiable in a citizen science search. We identified galaxy clusters which had previously been assigned as 'true lenses' as follows. We cross-matched the A–C grade lenses in the masterlens data base[2] (Moustakas 2012) with our object sample and retrieved those flagged as cluster-scale ('CLUST-GAL') from the data base. Since not all the objects in our object sample were also in the data base, we conducted a further visual inspection of the A–C graded candidates in our sample, flagging the cluster-scale lenses. Any objects identified by either method which had received a grade greater than the relevant cutoff ($\geq$B grade, or $\geq$C grade, as specified) were removed from the sample. These accounted for $\sim 25$ per cent of the lens systems in our sample. We show the corresponding ROC curves for the individual classifiers in this work with/without the clusters removed in Appendix A2. We find there is a small narrowing in the performance difference between the citizen science and neural network classifiers when the cluster-scale lenses are excluded, however, a combination of the classification method and use of lens subtraction means the citizen science classifier still outperforms the networks for our object sample.

[2] https://test.masterlens.org/

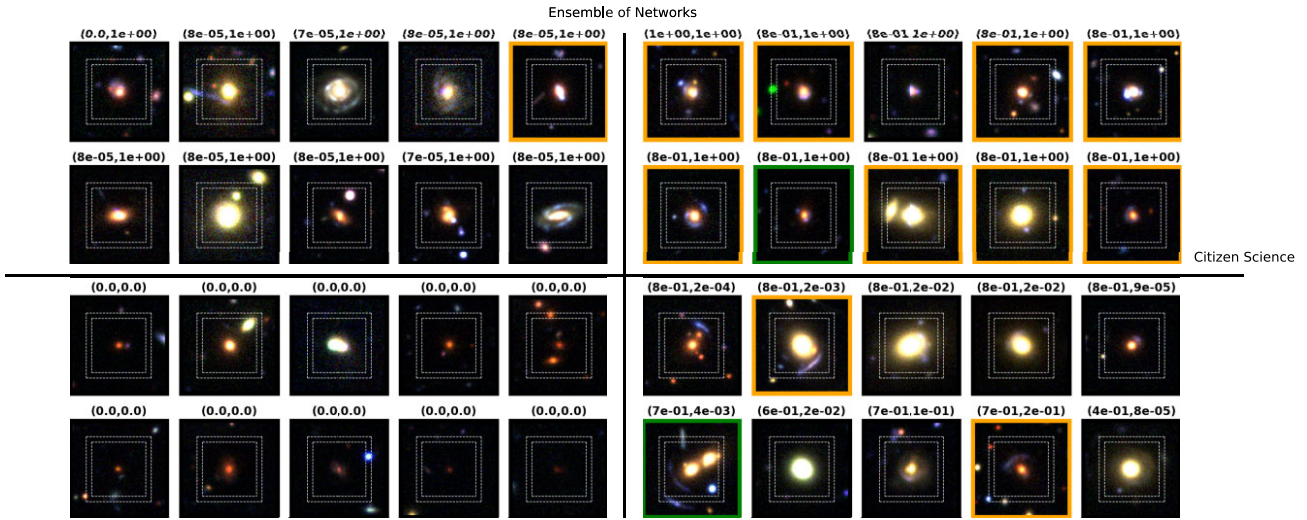

**Figure 6.** Example cutouts of high- and low-scoring objects from the citizen science search and the ensemble of five neural network classifiers. Objects with expert grades $1.5 < G < 2$ are outlined in orange, and those with $2 \leq G$ are outlined in green. The outer and inner dashed lines show the respective cutout sizes shown to the networks in Cañameras et al. (2021) and Shu et al. (2022), respectively, while the whole cutout was shown to the citizen scientists. The $(P_1, P_2)$ values above each cutout indicate the calibrated probabilities from the Citizen Science classifier and the calibrated posterior for the Network-only Ensemble, respectively.

Since removing cluster-scale lenses reduced our ground-truth sample size, we also investigated the inclusion of C-grade lenses as 'true lenses' in our ensemble. While a sizeable fraction may not be lenses in reality, this mimics the effect of increasing the sample size of lenses for the calibration, as may be available for a wider survey. We found that this improved the calibration for our classifiers; validation plots for this setup are shown in Appendix A3. In turn, we found this improved the performance of the classifier ensemble compared to the individual classifiers. Fig. 5(b) shows the ROC curves using A–C grade lenses as true lenses and with cluster-scale candidates removed. We found the ensemble of networks only (no CS) provided substantial improvement above the best individual network and performs better (in AUROC, FPR at 50 per cent completeness and true positive rate (TPR) at FPR $= 10^{-3}$) than the individual citizen science classifier. When citizen science was included, the ensemble further improved, increasing the completeness from 27 per cent (Network-only) to 40 per cent at FPR $= 10^{-3}$.

### 5.4 Expectations and implications for LSST

In Fig. 7, we compare the number of true and false positives expected for an LSST-like survey for an ensemble including and excluding the citizen science classifier. In these plots, we have scaled the total number of strong lenses to $10^5$ and used the true and false-positive rate functions stemming from the ensemble and individual classifiers applied to the HSC data in this work. We find that, when including citizen science in our sample, a 40 per cent complete sample can be achieved with 49 per cent purity, compared to 32 per cent purity for the best individual classifier. However, for higher completeness (towards the left of each plot), the sample would remain overwhelmingly false positives. We will further discuss the implications and mitigation for population-level analysis in a future work (Holloway et al. in prep). The difference is more substantial when the citizen science classifier is excluded, nearly doubling the purity to 28 per cent for a 40 per cent complete sample.

For such large samples, expert grading of all but the highest ranked candidates becomes intractable. It thus becomes even more important that the outputs of lens-finders are calibrated to allow statistical analysis of large samples of strong lenses including a known proportion of false positives. Fig. 8 shows the effective sample size of both the individual calibrated classifiers and the ensemble. It demonstrates the ensemble can retrieve a larger effective sample of lenses from the data, as well as a clear plateau, the 'knee' of which would be a useful initial starting point as a statistical sample of uncertain lenses. Having calibrated probabilities also allows comparison of objects inspected by different classifiers. This would allow rank-ordering across different samples of objects, not necessarily all seen by the same classifier which could be used for identifying candidates for follow-up and may be useful for selecting the forthcoming 4*MOST* sample (Collett et al. 2023).

Given that for high completeness, the number of expected false positives from LSST is still large, we discuss here further avenues for improving lens searches. Due to the huge data volume anticipated with forthcoming surveys, such alternatives must still minimize human intervention. Individual lens finding algorithms will continue to improve: the best network in Canameras et al. (2023) further improves upon that of Cañameras et al. (2021) used here (the former was not applied to the whole HSC survey) and suggest $TPR_0 \sim 60$ per cent could be achievable. With respect to ensemble classifiers, the large differences in the rank ordering of objects between classifiers (Fig. 4) suggest classifiers trained on a diverse range of training data find different types of non-lenses easier/more difficult to classify. This suggests that larger ensembles could offer

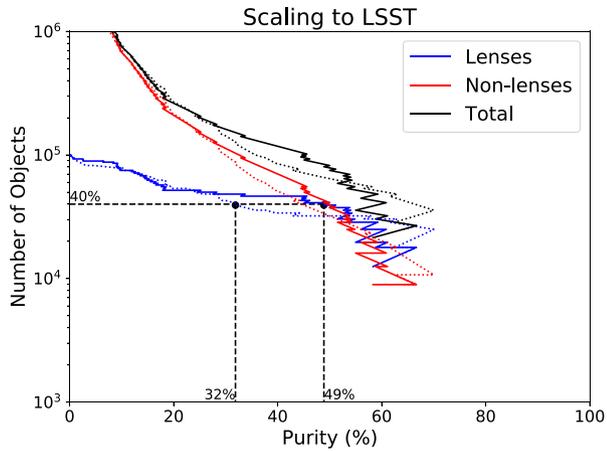

(a) An ensemble of 6 classifiers (solid lines) compared to only the citizen science classifier (dashed)

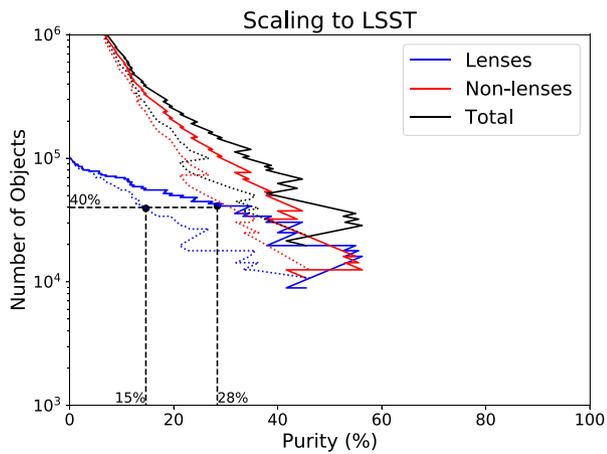

(b) An ensemble of the 5 neural networks (solid lines) compared to a single network (Neural Network 2, dashed)

**Figure 7.** A plot of the expected number of true and false positives for LSST; the total number of lenses has been fixed to $10^5$. These plots use the false-positive and true-positive rates as a function of the score threshold for the HSC classifiers used in this work (with A–B grade lenses as 'true' lenses), assuming the classifier performance in LSST matches those applied to HSC here. The curves have been cutoff when there were <5 lenses or non-lenses remaining in the test set, to reduce the effect of small-number statistics.

further improvement. We investigated this with the classifiers used in this work. We measured the AUROC, FPR$_{50}$ (false-positive rate at 50 per cent completeness) and TPR$_{-3}$ (completeness at an FPR of $10^{-3}$) as a function of number of classifiers in the ensemble, averaged over the combinations of available classifiers. We find the primary benefit in these metrics is achieved when adding $\geq 3$ classifiers, though the ensemble continued to improve up to the six used in this work. While very diverse, some aspects of the lens classifiers were similar (e.g. the lens model and non-lens sample used in Network 2 and 3) causing correlation between their scores. Although beyond the scope of this current work, it would be interesting to compare the performance of an ensemble of classifiers trained with non-overlapping samples of non-lenses with a single network trained on the whole data set, the benefit of an ensemble being that it should be able to mitigate the effects of any biases in an individual classifier.

The citizen science classifier (which used lens subtracted imaging) performed the best out of the classifiers used here however without significant increases in citizen numbers or object selection, it is likely unfeasible to apply citizen science to the whole LSST field (Marshall et al. 2016). Nevertheless, citizen science will still remain a key tool for lens-finding in this new era. The use of citizen-informed networks (i.e. active learning) such as applied to galaxy morphology classification (Walmsley et al. 2020) could be used instead to further improve automated classification. Similarly, using citizen science classifications as labelled training data for lens-finding neural networks would increase the size of a training set which used exclusively real images. Currently the number of known strong lenses is too small to train a network, which therefore typically rely on simulated lenses, however once data begins to arrive from wide-area surveys, citizen classifications of the real data could provide large volumes of labelled training data. Simultaneous fast lens modelling such as developed by Park et al. (2021) and Wagner-Carena et al. (2021) could also help improve current lens finding algorithms.

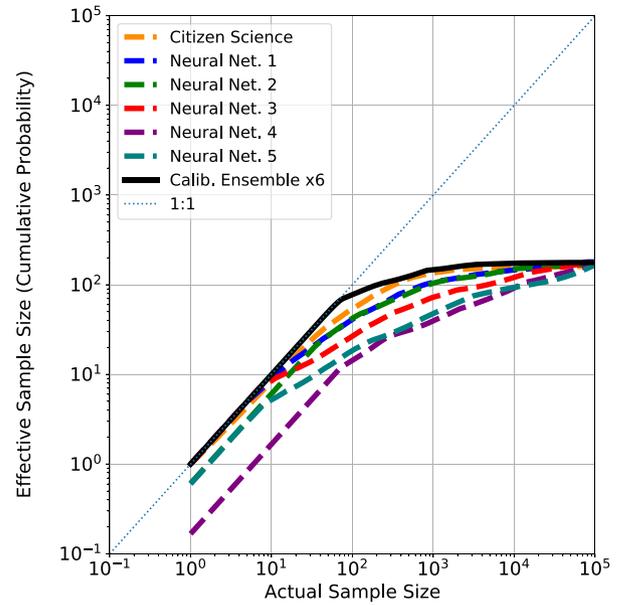

**Figure 8.** A plot of the effective sample size of the calibrated individual and ensemble classifiers presented above, as a function of total sample size. The effective sample size was determined by rank-ordering the objects in order of calibrated probability, then cumulatively summing them. The ensemble classifier used here is comprised of all six classifiers, and has been further recalibrated by isotonic regression after forming the ensemble. The ensemble classifier can produce the largest effective sample size out of all the classifiers.

A key component of the methodology described above is the presence of a ground truth from which to base the calibration. In this work, we compiled expert grades from a range of sources, in order to maximize the number of objects with assigned grades. While it may initially appear that having a known ground truth for a random sample of objects would be optimal for calibration, in practice the high-score regime is where calibration is most important and the calibration mapping changes the most rapidly. Citizen science could be used here to provide high-quality lens candidates to use as a ground truth for network calibration. For LSST, this ground truth could then be used for calibrating automated methods such as neural networks across the whole survey area.

## 6 CONCLUSION

This work had two fundamental aims: (1) to provide calibrated probabilities for a sample of galaxy cutouts that a given galaxy is a strong lens system and (2) to combine neural network and citizen science classifiers into an ensemble classifier, to maximize the purity of the resulting sample, without compromising on completeness. We used six classifiers (one citizen science search and five neural networks) previously applied to HSC data, chosen as a proxy for the forthcoming LSST. Having achieved these aims, our conclusions are as follows:

(i) It is possible to post-process the outputs of a given lens classifier to produce accurate calibrated probabilities that a given classified object is a lens. The original outputs of such classifiers are not *a priori* calibrated probabilities, and should not be treated as such in a statistical analysis.

(ii) There is very little correlation between classifiers of the scores of non-lenses. This can be used advantageously to help remove false positives, as different classifiers find certain systems easier/more difficult to classify.

(iii) An ensemble classifier can provide improved classification above its constituent components. For an FPR of $10^{-3}$, an ensemble classifier can increase the completeness of the resultant sample from 34 per cent to 46 per cent.

(iv) Given $\sim 10^5$ strong lenses in LSST, further improvements in scalable lens finding methods will be needed in order to achieve completenesses > 50 per cent without significant contamination by false positives.

## ACKNOWLEDGEMENTS

We thank the anomymous referee for their very useful comments and suggestions for this paper. PH would like to thank the KIPAC Strong Lensing group for very useful discussions and Alessandro Sonnenfeld and Yiping Shu for providing data for this project.

PH acknowledges funding from the Science and Technology Facilities Council, Grant Code ST/W507726/1 including for a long-term attachment to the SLAC National Accelerator Laboratory. The work of PJM was supported by the U.S. Department of Energy under contract number DE-AC02-76SF00515. AV acknowledges support from the Science and Technology Facilities Council, grant code ST/S006168/1 and ST/X00127X/1. This research was supported in part by the Max Planck Society, and by the Excellence Cluster ORIGINS which was funded by the Deutsche Forschungsgemeinschaft (DFG, German Research Foundation) under Germany's Excellence Strategy – EXC – 2094 – 390783311. This project received funding from the European Research Council (ERC) under the European Unions Horizon 2020 research and innovation programme (LENSNOVA: grant agreement no. 771776). ATJ was supported by the Program Riset Unggulan Pusat dan Pusat Penelitian (RU3P) of LPIT Institut Teknologi Bandung 2023. This work was supported by JSPS KAKENHI grant number JP20K14511.

## DATA AVAILABILITY

Data underlying this article were provided by the sources listed in Section 2. Requests for such data should be made to the corresponding authors.

## REFERENCES

Aihara H. et al., 2018, PASJ, 70, S8
Aihara H. et al., 2019, PASJ, 71, 114
Aihara H. et al., 2022, PASJ, 74, 247
Akeson R. et al., 2019, preprint (arXiv:1902.05569)
Andika I. T. et al., 2023, A&A, 678, A103
Bolton A. S., Burles S., Koopmans L. V. E., Treu T., Moustakas L. A., 2006, ApJ, 638, 703
Bolton A. S., Burles S., Koopmans L. V. E., Treu T., Gavazzi R., Moustakas L. A., Wayth R., Schlegel D. J., 2008, ApJ, 682, 964
Browne I. W. A. et al., 2003, MNRAS, 341, 13
Brownstein J. R. et al., 2012, ApJ, 744, 41
Cañameras R. et al., 2021, A&A, 653, L6
Canameras R. et al., 2023, preprint (arXiv:2306.03136)
Collett T. E., 2015, ApJ, 811, 20
Collett T. E. et al., 2023, The Messenger, 190, 49
Euclid Collaboration, 2022, A&A, 662, A112
Faure C. et al., 2008, ApJS, 176, 19
Garvin E. O., Kruk S., Cornen C., Bhatawdekar R., Cañameras R., Merín B., 2022, A&A, 667, A141
Geach J. E. et al., 2015, MNRAS, 452, 502
Geng S., Cao S., Liu Y., Liu T., Biesiada M., Lian Y., 2021, MNRAS, 503, 1319
Hartley P., Flamary R., Jackson N., Tagore A. S., Metcalf R. B., 2017, MNRAS, 471, 3378
He Z. et al., 2020, MNRAS, 497, 556
Holloway P., Verma A., Marshall P. J., More A., Tecza M., 2023, MNRAS, 525, 2341
Holwerda B. W. et al., 2019, AJ, 158, 103
Ivezić Ž. et al., 2019, ApJ, 873, 111
Jackson N., 2008, MNRAS, 389, 1311
Jacobs C., Glazebrook K., Collett T., More A., McCarthy C., 2017, MNRAS, 471, 167
Jacobs C. et al., 2019a, ApJS, 243, 17
Jacobs C. et al., 2019b, MNRAS, 484, 5330
Jaelani A. T., More A., Wong K. C., Inoue K. T., Chao D. C. Y., Premadi P. W., Cañameras R., 2023, preprint (arXiv:2312.07333)
Knabel S. et al., 2020, AJ, 160, 223
Lanusse F., Ma Q., Li N., Collett T. E., Li C.-L., Ravanbakhsh S., Mandelbaum R., Póczos B., 2018, MNRAS, 473, 3895
Li R. et al., 2021, ApJ, 923, 16
Lintott C. J. et al., 2008, MNRAS, 389, 1179
Marshall P. J. et al., 2016, MNRAS, 455, 1171
More A. et al., 2016, MNRAS, 455, 1191
Moustakas L., 2012, The Master Lens Database and The Orphan Lenses Project, HST Proposal ID 12833. Cycle 20, (accessed 8 September 2023)
Myers S. T. et al., 2003, MNRAS, 341, 1
Park J. W., Wagner-Carena S., Birrer S., Marshall P. J., Lin J. Y.-Y., Roodman A., LSST Dark Energy Science Collaboration, 2021, ApJ, 910, 39
Pascale M. et al., 2022, ApJ, 938, L6
Pawase R. S., Courbin F., Faure C., Kokotanekova R., Meylan G., 2014, MNRAS, 439, 3392
Petrillo C. E. et al., 2017, MNRAS, 472, 1129
Petrillo C. E. et al., 2019, MNRAS, 484, 3879
Platt J., 2000, in Smola A.J., Bartlett P., Schölkopf B., Schuurmans D., eds, Advances Large-Margin Classifiers, Vol. 10, Neural Information Processing series. MIT Press, MA, USA
Powell D. M., Vegetti S., McKean J. P., Spingola C., Stacey H. R., Fassnacht C. D., 2022, MNRAS, 516, 1808
Powell D. M., Vegetti S., McKean J. P., White S. D. M., Ferreira E. G. M., May S., Spingola C., 2023, MNRAS, 524, L84
Rojas K. et al., 2022, A&A, 668, A73
Rojas K. et al., 2023, MNRAS, 523, 4413
Sahu K. C. et al., 2022, ApJ, 933, 83
Savary E. et al., 2022, A&A, 666, A1
Schaefer C., Geiger M., Kuntzer T., Kneib J. P., 2018, A&A, 611, A2
Seidel G., Bartelmann M., 2007, A&A, 472, 341
Shu Y., Cañameras R., Schuldt S., Suyu S. H., Taubenberger S., Inoue K. T., Jaelani A. T., 2022, A&A, 662, A4
Sonnenfeld A. et al., 2018, PASJ, 70, S29

Sonnenfeld A., Jaelani A. T., Chan J., More A., Suyu S. H., Wong K. C., Oguri M., Lee C.-H., 2019, A&A, 630, A71
Sonnenfeld A. et al., 2020, A&A, 642, A148
Spergel D. et al., 2015, preprint (arXiv:1503.03757)
Sugiyama M., Suzuki T., Nakajima S., Kashima H., Bünau P. V., Kawanabe M., 2008, Ann. Inst. Stat. Math., 60, 699
Thuruthipilly H., Zadrozny A., Pollo A., Biesiada M., 2022, A&A, 664, A4
Wagner-Carena S., Park J. W., Birrer S., Marshall P. J., Roodman A., Wechsler R. H., LSST Dark Energy Science Collaboration, 2021, ApJ, 909, 187
Walmsley M. et al., 2020, MNRAS, 491, 1554
Zadrozny B., Elkan C., 2002, Proceedings of the ACM SIGKDD International Conference on Knowledge Discovery and Data Mining. ACM Digital Library, New York, NY, USA, p. 694

## APPENDIX A:

### A1 ROC curves for different ensemble methods

In Fig. A1, we show the ROC curves for an ensemble classifier, combining the classifiers using the methods described in Section 3.3. We find that, in the low false-positive region, the Independent Bayesian methods provide the greatest completeness.

### A2 ROC curves using A–B grade candidates with/without clusters

In Fig. A2, we show the ROC curves for the individual lens classifiers when altering the ground truth for 'true lenses'. We observe a small narrowing between the performance of the neural networks and that of the citizen science classifier when cluster-scale lenses are excluded.

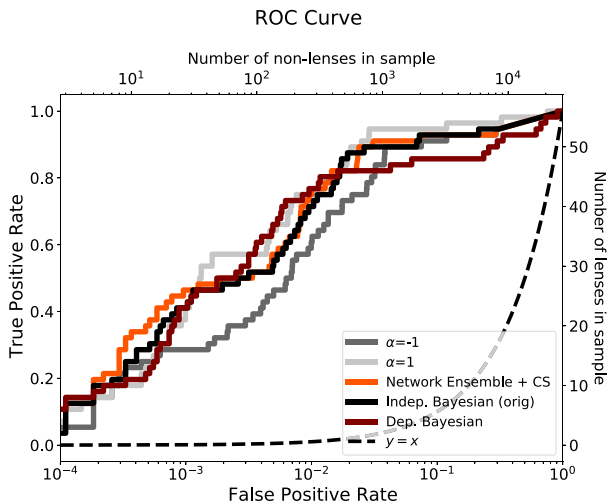

**Figure A1.** ROC Curves for a selection of different classifier-combination methods. The best-performing methods at the low-false-positive end were the Independent Bayesian methods (with/without generating a network-only ensemble first), which are used in the rest of the paper. The ROC curve for a random classifier is shown as the dashed curve ($y = x$). The $\alpha$ labels refer to the generalized mean combination from Section 3.3.1.

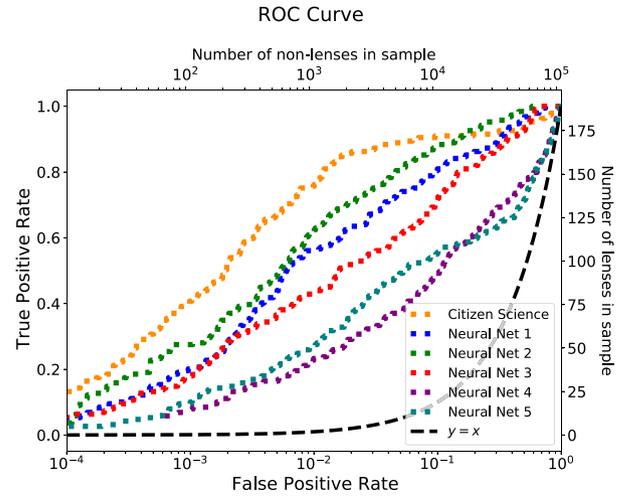

(a) Including A-B grade lenses as 'true lenses', without excluding cluster-scale candidates.

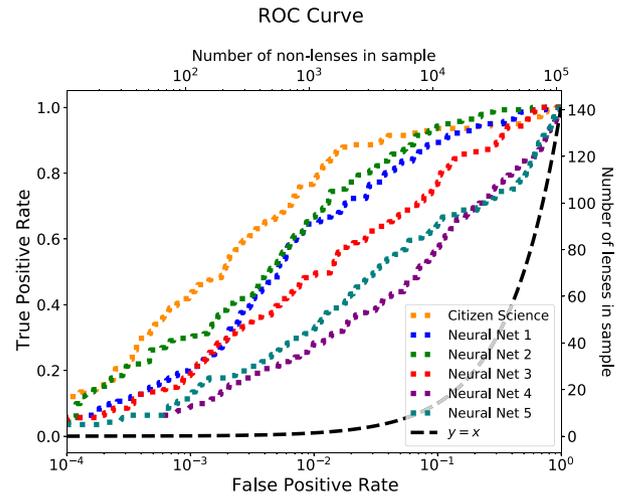

(b) Including A-B grade lenses as 'true lenses', and excluding cluster-scale candidates.

**Figure A2.** ROC curve for the individual lens classifiers, applied to the whole cross-matched data set. The ROC curve for a random classifier is shown as the dashed curve ($y = x$). Note the $y$-axes are linear, while the $x$-axes are logarithmic.

### A3 Calibration validation curves with A–C grade candidates, excluding clusters

In Fig. A3, we show the validation plots for each calibration method, when treating A–C grade lenses (but not cluster-scale systems) as 'true lenses'. Overall we see improved calibration accurately reaching higher calibrated probabilities.

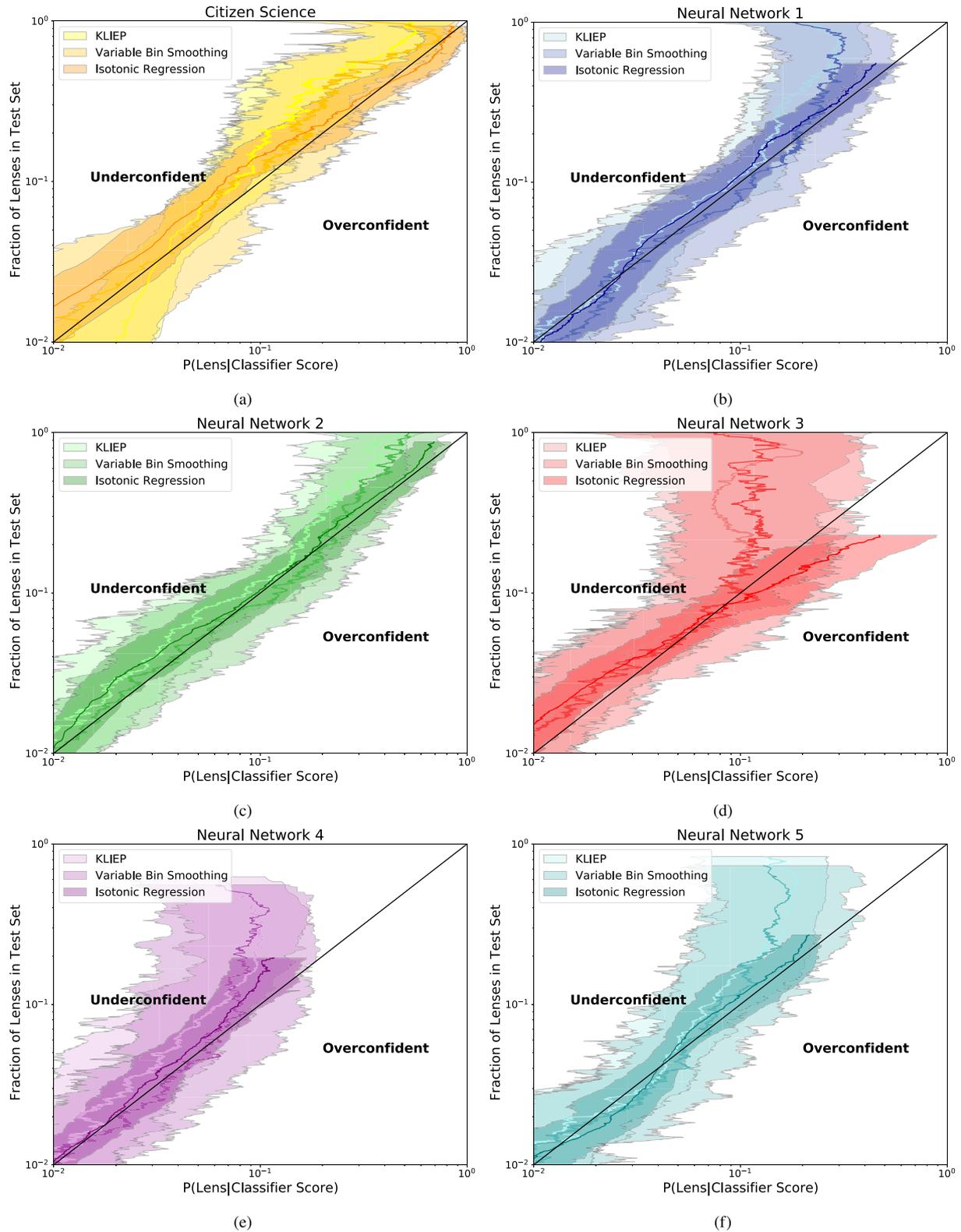

**Figure A3.** Validation of the calibration curves, applied to a separate test set of graded images. A–C grade lenses are counted as 'true lenses' for these plots, and cluster-scale candidates receiving A–C grades have been excluded.

This paper has been typeset from a T<sub>E</sub>X/L<sup>A</sup>T<sub>E</sub>X file prepared by the author.